# Millimeter-scale exfoliation of hBN with tunable flake thickness

Amy S. Mckeown-Green[1], Helen J. Zeng[1], Ashley P. Saunders[1], Jiayi Li[1], Jiaojian Shi[2,3], Yuejun Shen[2,3], Feng Pan[2], Jenny Hu[4], Jennifer A. Dionne[2,5], Tony F. Heinz[4,6], Stephen Wu[7], Fan Zheng[8], Fang Liu[1*]

1. Department of Chemistry, Stanford University, Stanford, CA 94305, USA
2. Department of Materials Science and Engineering, Stanford University, Stanford, CA 94305, USA
3. Stanford Institute for Materials and Energy Sciences, SLAC National Accelerator Laboratory, Menlo Park, CA 94025, USA
4. Department of Applied Physics, Stanford University, Stanford, CA 94305, USA
5. Department of Radiology, Stanford University, Stanford, CA 94305, United States
6. Stanford PULSE Institute, SLAC National Accelerator Laboratory, Menlo Park, CA 94025, USA
7. Department of Electrical and Computer Engineering, University of Rochester, Rochester, New York 14627, USA
8. School of Physical Science and Technology, ShanghaiTech University, Shanghai 201210, China

*Corresponding author: fliu10@stanford.edu

## Abstract

As a two-dimensional (2D) dielectric material, hexagonal boron nitride (hBN) is in high demand for applications in photonics, nonlinear optics, and nanoelectronics. Unfortunately, the high-throughput preparation of macroscopic-scale, high-quality hBN flakes with controlled thickness is an ongoing challenge, limiting device fabrication and technological integration. Here, we present a metal thin-film exfoliation method to prepare hBN flakes with millimeter-scale dimension, near-unity yields, and tunable flake thickness distribution from 1-7 layers, a substantial improvement over scotch tape exfoliation. The single crystallinity and high quality of the exfoliated hBN are demonstrated with optical microscopy, atomic force microscopy, Raman spectroscopy, and second harmonic generation. We further explore a possible mechanism for the effectiveness and selectivity based on thin-film residual stress measurements, density functional theory calculations, and transmission electron microscopy imaging of the deposited metal films. We find that the magnitude of the residual tensile stress induced by thin film deposition plays a key role in determining exfoliated flake thickness in a manner which closely resembles 3D semiconductor spalling. Lastly, we demonstrate that our exfoliated, large-area hBN flakes can be readily incorporated as encapsulating layers for other 2D monolayers. Altogether, this method brings us one step closer to the high throughput, mass production of hBN-based 2D photonic, optoelectronic, and quantum devices.

## Introduction

Within the diverse cornucopia of two-dimensional (2D) materials, hexagonal boron nitride (hBN) stands out as an atomically thin dielectric with a wide variety of applications in photonics,[1,2] non-linear optics,[3–5] and nanoelectronics.[6,7] Its suitability for these applications is supported by its wide band gap (~6 eV) and diverse thermal, optical, and electronic properties.[8] When isolated as a monolayer, hBN has been used in van der Waals heterojunctions as a dielectric tunneling barrier to tune the electronic coupling between adjacent layers.[9] Monolayer hBN also has an exceptionally large, in-plane thermal conductivity and is second only to diamond in a measurement of thermal conductivity per unit weight.[8] As such, it has great potential in nanoelectronics as a heat dissipation layer.[10,11] Most recently, monolayer hBN has been used to observe moiré engineered ferroelectricity in a parallel-stacked configuration.[12] Thicker hBN flakes, ranging from a few to tens of layers thick, are also highly desired for various applications. When used as an encapsulating layer in 2D devices, these multilayer hBN flakes protect air-sensitive 2D materials from ambient conditions, maintaining their exotic electronic, ferromagnetic, or quantum properties.[8,13,14] Multilayer hBN flakes are also broadly used as a dielectric in 2D devices, such as in top-gate field effect transistors (FETs), where they minimize the presence of interfacial trap states and optimize device performance.[15] Lastly, multilayer hBN can be integrated into high-quality factor waveguides, resonators, and metasurfaces to induce a strong, non-linear optical response without visible optical losses.[2,16] To address these diverse applications, the development of a large-scale, high-yield, and *thickness-selective* method for hBN flake preparation is of key significance to the field.[17]

Despite its significance, the large-area production of hBN with precise thickness control is still an ongoing challenge. Top-down methods, such as Scotch tape exfoliation, have historically been used to prepare high-quality, single-crystalline hBN flakes.[17] Unfortunately, these flakes typically suffer from very small lateral dimensions (up to tens of µm), uncontrolled thickness (monolayer to micrometers), and extremely low overall yield.[18,19] This is in part due to the partial-ionic interlayer bonding present in hBN which makes mechanical exfoliation more challenging compared to other 2D materials, such as graphene.[20,21] These tape-based exfoliation methods are also stochastic, making flake identification and characterization a time-consuming process. In contrast, recent advancements in bottom-up techniques, such as chemical vapor deposition (CVD), have enabled the growth of single-crystalline, monolayer hBN on molten or precisely faceted substrates at the wafer scale.[22–24] However, these CVD films still suffer from relatively high defect concentrations, and precise control of layer thickness beyond monolayer has not yet been widely realized.[19,25]

Exfoliation via metal thin films has proved to be a promising avenue for obtaining large-area, thin-layer flakes for a variety of 2D materials.[26–29] For transition metal dichalcogenides (TMDCs), the selective exfoliation of millimeter-to-centimeter scale monolayers has been achieved via direct gold deposition,[26] exfoliation onto a gold substrate,[30] and use of a gold tape created by template stripping.[31,32] The mechanism underlying this selectivity is the quasi-bonding interaction between gold and the top monolayer, which is strong enough to overcome the native interlayer van der Waals force.[30–32] Similarly, direct metal deposition has been used to exfoliate graphene flakes, where exfoliated flake thickness was affected by different types of metal and the resulting graphene-metal binding energies.[29] In comparison to TMDCs and graphene, the thickness selective exfoliation of large-area hBN is far less developed. Recent work has demonstrated the exfoliation of multilayer hBN flakes via direct gold deposition. However, flake sizes were limited to < 75 μm, and selective thickness control was not demonstrated.[28] A macroscopic exfoliation technique with near unity yield and simultaneous control over exfoliated flake thickness is therefore a key area for development.

Here, we present a metal-assisted exfoliation technique optimized to prepare millimeter-scale hBN flakes, employing six different metals to achieve tunable flake thicknesses ranging from 1-7 layers, reaching a significantly higher selectivity compared with scotch tape exfoliation. We demonstrate the high optical and crystalline quality of the exfoliated hBN via optical microscopy, atomic force microscopy (AFM), Raman spectroscopy, and second harmonic generation (SHG). To understand the mechanism underlying the observed flake-thickness selectivity, we investigate the crystallinity, mechanical properties, and binding energies of the deposited metal thin films using transmission electron microscopy (TEM), thin film profilometry, and density functional theory (DFT) calculations. Our results suggest that residual stress resulting from direct metal deposition is a dominant factor for determining flake thickness. The access to high-quality, macroscopic, and single-crystalline hBN flakes with tunable thickness provided by this technique has significant implications for the future commercialization and mass production of hBN-based 2D devices.

## Results and Discussion

A schematic of this metal-assisted exfoliation method is shown in Fig 1a. First, the metal of choice is deposited on the flat surface of a freshly cleaved bulk hBN crystal via electron beam evaporation. Next, a polyvinylpyrrolidone (PVP) protective layer is spin-coated on the metal surface to guard against contamination from the thermal release tape (TRT).[31] The TRT is then used as an adhesive handle to exfoliate thin layer hBN from the bulk crystal and transfer it to the desired substrate.

The TRT is released in water at 97°C. The use of an aqueous environment facilitates PVP dissolution, allowing for an even TRT release, thereby enhancing the yield of thin layer hBN. Following additional water rinses to remove residual PVP, the metal layer is chemically etched, yielding thin layer hBN (see Methods Section). Fig. 1b and 1c show optical images of a prototypical millimeter-scale thin layer hBN flake as it appeared on the metal/TRT and on a 90 nm $SiO_2$/Si substrate. This method is highly reproducible, and flakes of similar or larger sizes were consistently exfoliated (See Fig. S1 and S2).

We demonstrate this metal-assisted exfoliation procedure with six different metals: gold (Au), palladium (Pd), indium (In), copper (Cu), nickel (Ni), and titanium (Ti). Large-area hBN flakes with a thickness of less than 25-layers were consistently exfoliated (see Fig. S2).

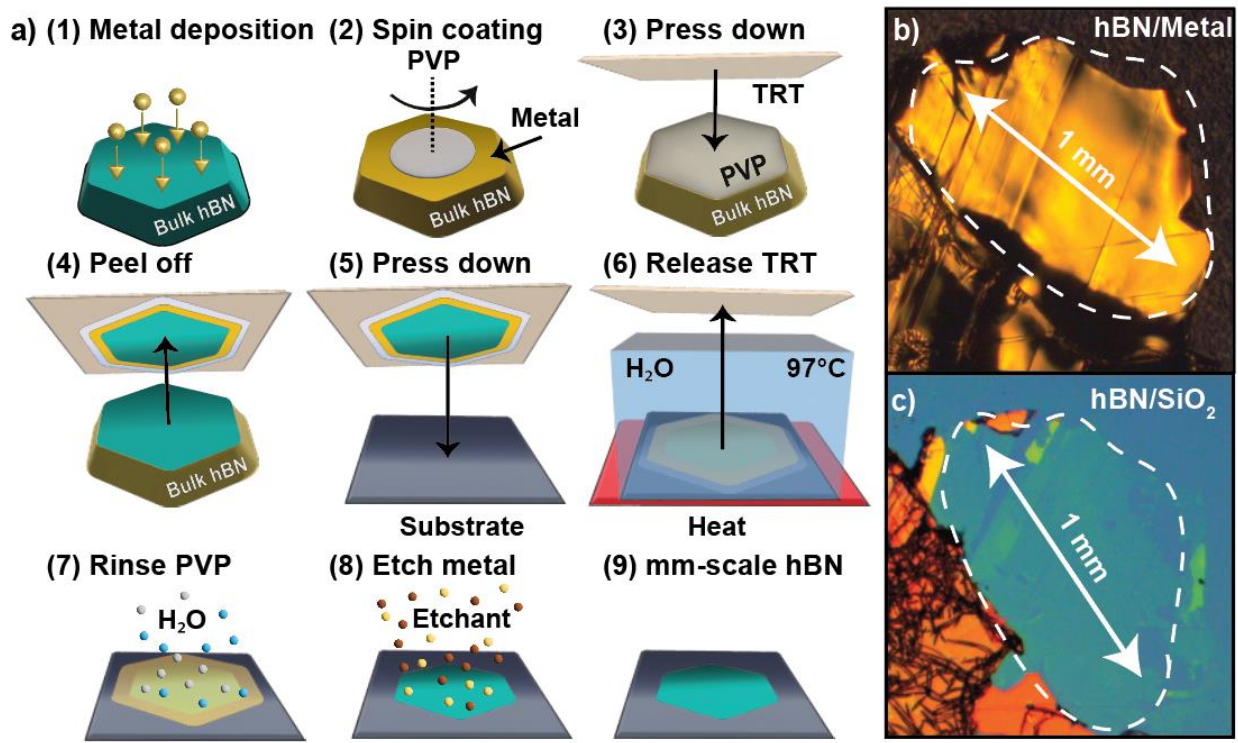

**Figure 1 (a)** Schematics of exfoliation method: (1) metal deposition on bulk hBN crystal; (2) spin coating a layer of PVP onto the metal surface; (3) press down and (4) peel off the layer with thermal release tape (TRT); (5) transferring to a substrate; (6) TRT removal via heating in water; (7) rinsing of remaining PVP; (8) Etching of the metal; (9) Obtaining millimeter scale thin-layer hBN. **(b)** Optical image of thin layer hBN on gold. **(c)** Optical image of the same thin-layer hBN transferred on a 285 nm SiO$_2$/Si substrate.

**High optical and crystalline quality**

To examine the optical quality of hBN flakes exfoliated with different metals, we obtained Raman spectra of the 2D, in-plane $E_{2g}$ phonon mode for hBN flakes exfoliated with Au, Pd, In, Ni, Ti, and Cu (Fig. 2a). While substrate-induced strain effects cause small shifts in central frequency, the $E_{2g}$ peak positions and full width at half maxima (FWHM) are in good agreement with those obtained for Scotch tape exfoliated flakes prepared from the same hBN bulk crystals. (See Supporting Information).

Next, a ~0.2 mm large monolayer flake, exfoliated with Au, was characterized via a SHG polarization scan. The corresponding scan shown in Fig 2b confirms the flake's monolayer thickness, single crystallinity, and high optical quality.[19] Fig. 2c shows a topographical AFM map on the edge of the same monolayer flake. The corresponding line profile in Fig. 2d demonstrates a sharp crystallographic edge and confirms monolayer flake thickness.[27]

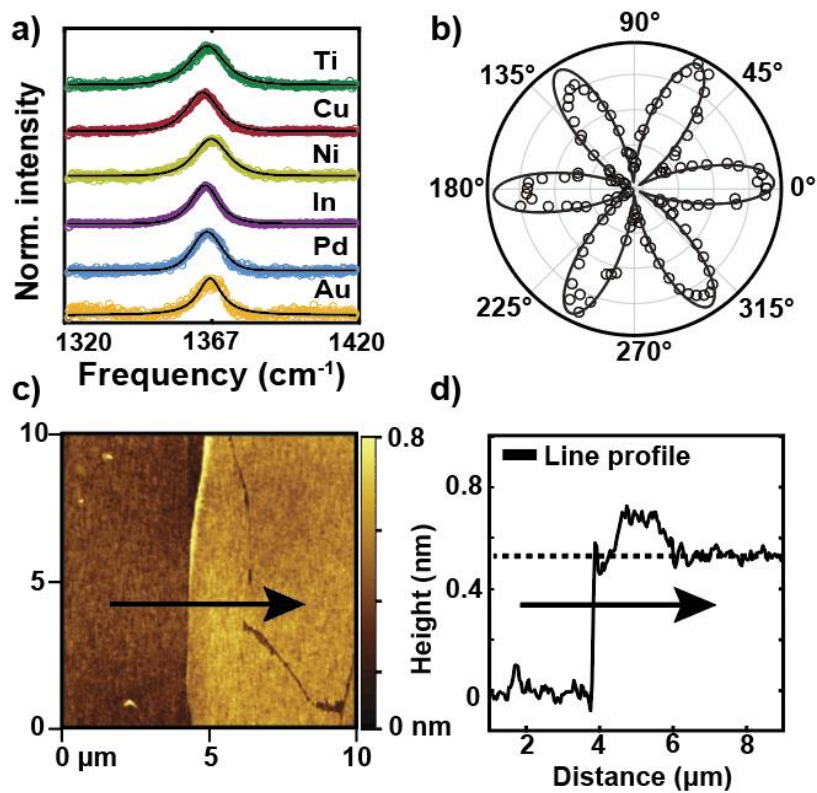

**Figure 2. (a)** Raman spectra of the $E_{2g}$ peak for hBN flakes with thickness ranging from 3-5 layers exfoliated with different metals (*11 layers for Cu). (**b**) SHG polarization scan of monolayer hBN. (**c**) Topographical AFM image of monolayer hBN. (**d**) Line profile showing step-height corresponding to monolayer hBN featured in panel (**c**).

**Metal-specific flake thickness selectivity**

We determined the thickness-dependent distributions of hBN flakes exfoliated with each metal, using the corresponding flake areas and thicknesses measured via optical microscopy and AFM. The results are shown in Fig. 3a. In, Pd, and Au show distinct flake thickness distributions with maximum exfoliation probabilities of 1-layer, 3-layers, and 7-layers respectively. We observe the same selectivity in the average and maximum flake size distributions (see Supporting Information). With regards to flake size, 3-layer (Pd-exfoliated) and 7-layer (Au-exfoliated) hBN flakes were found to have maximum lateral dimensions at the sub-millimeter to millimeter scale. hBN flakes exfoliated with In were found to have maximum flake sizes on the order of 6400 μm². While In-exfoliated hBN flakes were smaller than Au- or Pd-exfoliated flakes, it should be noted that their length scale still far exceeds monolayer hBN sizes typically achieved with Scotch tape exfoliation.[33]

Beyond these flake area distributions, we determined the exfoliation yield for each metal by dividing the area of the exfoliated thin-layer hBN (< 25 layers) by the total, flat surface area of the parent bulk crystal (Equation 1).

$$Percent\ Yield\ (\%) = \frac{Total\ area\ of\ thin-layer\ hBN}{Area\ of\ flat\ surface\ on\ parent\ bulk\ crystal} \times 100\% \quad (1)$$

Fig. 3b shows the thin-layer hBN yields from repeated exfoliations of the same group of bulk hBN crystals with differing size and surface quality. Au-exfoliations were observed to have high thin-layer hBN yields of up to 95%. In contrast, Pd- and In-exfoliations were observed to have maximum yields of 65% and 43% respectively. For each metal, we find that thin-layer yield is highly dependent on the initial surface quality of the hBN bulk crystals. We observe that, while the bulk hBN crystals could have lateral dimensions up to 5 mm, they often consist of ≥3 large crystallographic domains. The variation in domain size and bulk crystal quality were key contributors to the distribution in percent yields obtained for each metal (Fig. 3b). The variation in yield between metals is partially determined by metal-specific differences in their mechanical and chemical properties, including their response to stress/deformation, post-deposition oxidation, and unique chemical etching conditions which can promote flake delamination.

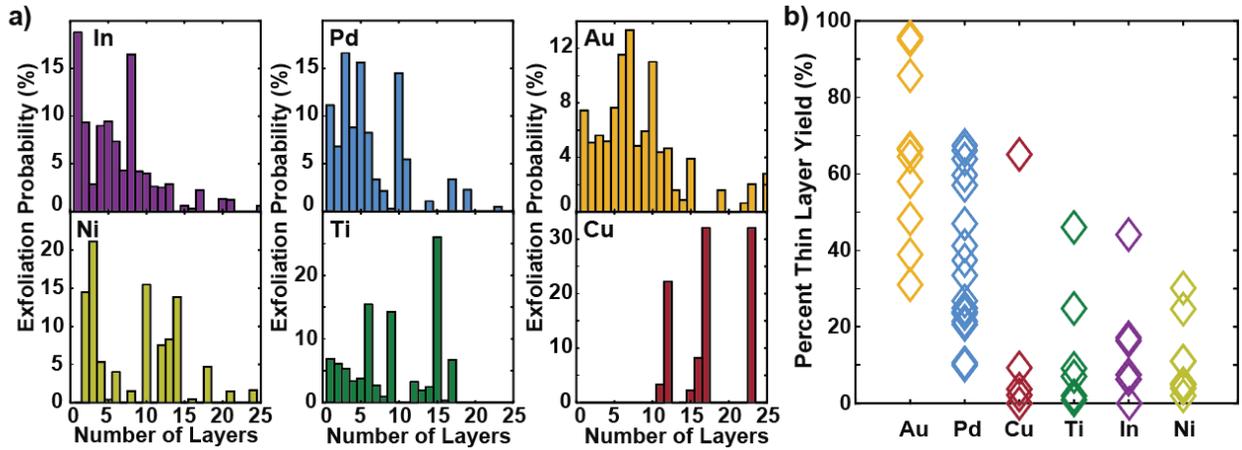

**Figure 3.** (**a**) Thickness distribution of thin hBN flakes exfoliated with In, Pd, Au, Ni, Ti, and Cu. (**b**) Metal-assisted exfoliation yield of thin layer hBN using the same six metals. Near-unity yields were only observed for pristine, flat, single-domain bulk crystals.

The flake thickness distributions for Ni and Ti exfoliation are less conclusive, and Cu is only observed to exfoliate flakes 11 layers or thicker. Cu-, Ni-, and Ti-exfoliated hBN flakes had maximum areas > 9000 μm$^2$, with peak exfoliation yields between 25-65% (Fig 3b). In addition to the aforementioned metals, this exfoliation method was performed with silver (Ag) and aluminum (Al), both with negligible thin-layer hBN yields. Interestingly, Ag is known to exfoliate TMDCs and other 2D materials.[34,35] We posit that this negligible thin-layer hBN yield is due to delamination during the etching process, leading to a significant loss of exfoliated hBN flakes. Further optimization of the etching conditions might improve exfoliation outcomes for Ag and Al, providing a lower cost alternative to Au and Pd.

**Exfoliation mechanism**

To obtain a deeper understanding of the exfoliation mechanism underlying flake thickness selectivity, we investigate the nature of the metal-bulk hBN interfacial interactions through TEM, DFT calculations, and stress measurements. We begin by examining the alignment of the hBN and metal lattices as interfacial epitaxy is a suggested mechanism for flake thickness selectivity in similar metal-assisted exfoliations.[36] In this epitaxial picture, the (111) surface of the metal and hBN lattices are aligned.[37] The resulting lattice mismatch induces epitaxial strain, enabling the selective cleavage of the top monolayer from the bulk crystal.[36] To determine whether such hBN/metal epitaxy was present in our samples, we employ TEM and selected area electron diffraction (SAED) to characterize both the degree of metal thin-film crystallinity and the overall alignment of the hBN and metal lattices.

Fig 4a shows a representative SAED pattern of a 30 nm gold thin film deposited on an hBN flake (< 20 nm thick). The highly localized, bright single crystalline diffraction spots correspond to the $\{10\bar{1}0\}$ and $\{11\bar{2}0\}$ planes of the hBN flake.[38] In comparison, the gold thin film exhibits polycrystalline diffraction rings which correspond mainly to the {220} planes and higher order reflections. Being the lowest energy surface, the Au {111} planes are favored during deposition. The majority of grains are observed to align with the Au {111} planes parallel to the $(000\bar{1})$ surface of the hBN crystal. This is also confirmed via X-ray diffraction (see Supporting Information).[39] However, the azimuthally uniform Au diffraction rings, particularly the {220} family of planes, indicate that there is no epitaxial alignment between the hBN and gold lattice. We therefore conclude that epitaxial growth is not a dominant factor in this work. However, pseudo-epitaxy between face-centered cubic metals and hBN has been observed under different deposition conditions and may be an interesting avenue to explore in future experiments.[37]

To further investigate the nature of metal-hBN interface, we compute the binding strengths (interfacial energies) between the different metal (111) surfaces and hBN crystal surface using first-principles density functional theory (DFT). The binding strengths for each metal of interest, defined as $|\gamma| = |E_{\text{metal-hBN}} - (E_{\text{metal}} + E_{\text{hBN}})|$ per BN formula unit, are shown in Fig. 4b. The specific numerical values can be found in Table S2. We observe that In, Pd, and Au exhibit hBN binding energies of similar or higher magnitude to hBN's interlayer binding energy (dashed line in Fig. 4b), enabling the efficient exfoliation of hBN layers from the bulk crystal.

In the previous reports on exfoliation of the graphene, higher metal binding energies were found to be correlated with thicker graphene flakes.[29] However, we do not observe a monotonic trend between hBN thickness selectivity and metal binding energies. Pd exhibits a relatively stronger hBN binding strength than either Au and In. Conversely, it exhibits a flake thickness selectivity which is in between that observed for Au and In. This suggests that the interfacial binding energy alone cannot solely predict the exfoliated-hBN-thickness dependence. We propose that this difference in graphene and hBN metal-assisted exfoliations lies in their unique electronic and mechanical properties, which might alter the relative role of the binding energies in determining exfoliated flake thickness. For example, hBN's increased mechanical stiffness compared to graphene is known to amplify strain propagation from thin films, which may make strain a more dominating factor in hBN exfoliations.[40]

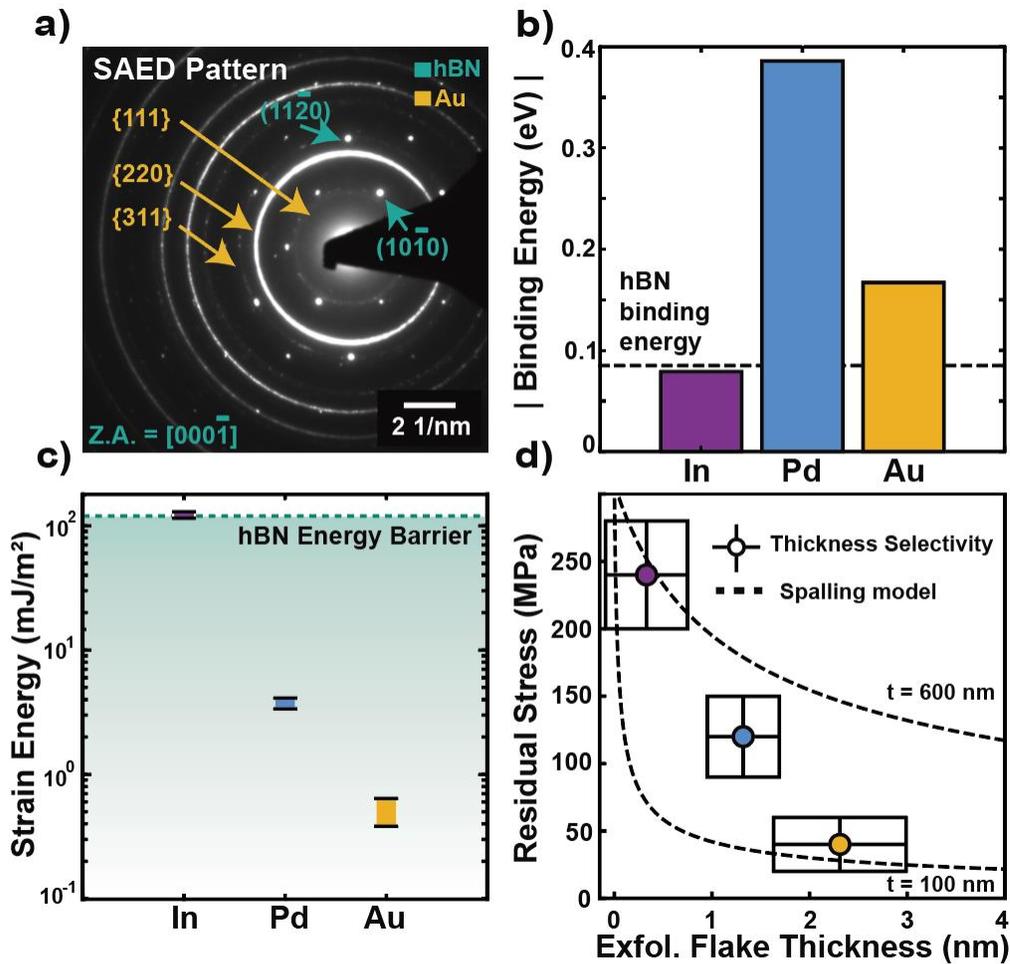

Figure 4. (a) Selected area electron diffraction pattern absence of epitaxy between the Au lattice (gold) and hBN (green). Gold arrows point to prominent or nearly-absent (e.g. {111}) polycrystalline diffraction rings from the Au thin film. Green arrows point to single crystalline diffraction spots from hBN. (b) Metal-hBN binding energies per BN formula unit for In, Pd, and Au thin-films calculated via DFT. (c) Calculated tensile strain energy applied laterally by metal thin-film deposition compared to the hBN energy barrier to shear motion. (d) hBN flake thickness as a function of tensile residual stress due to thin-film deposition accompanied by an analytical spalling model for two starting bulk crystal thicknesses (100 nm and 600 nm). Error bars represent the uncertainty in metal-specific, flake thickness distributions due to noise in the AFM measurement.

To conclude our investigation of the hBN-metal interface, we calculate the accumulated strain energy (Fig 4c) in the stressor thin-films and then compare them to the energy threshold of hBN's interlayer shear motion (i.e. how easily hBN layers can slide across one another).[41] To calculate this interfacial strain energy, we first determine the degree of residual stress in the evaporated metal thin films. Residual stress is widely observed in metal thin films as a result of their growth and microstructural evolution (i.e. how grains form, coalesce, and grow) during deposition.[42] To evaluate this residual stress in our metal thin film depositions, we measured the change in curvature of a thin, glass coverslip resulting from metal deposition using profilometry.[40,43,44] As shown in Fig. 4d, we observe that the deposition of Au, Pd, and In thin films induced tensile stress, which is as expected for these high mobility metals.[29] Overall, the In thin film exhibited the highest magnitude of tensile stress, followed by Pd and Au. Noticeably, this trend in residual stress closely matches the observed trend in flake thickness selectivity across the three metals (shown in Fig. 3a), with higher residual stress corresponding to generally thinner hBN flakes.

We calculate the total strain energy ($U_{strain}$) of the metal thin-films using Equation 2, with each metal's characteristic Young's modulus ($Y_{film}$), Poisson's ratio ($v_{film}$), and experimentally measured residual stress ($\sigma_{film}$).[45–47]

$$U_{strain} = \frac{(1-v_{film})}{2Y_{film}} t_{film} \sigma_{film}^2 \tag{2}$$

Fig. 4c shows the calculated strain energies for the Au, Pd, and In thin films. The dashed line represents hBN's energy barrier to interlayer motion, previously determined using molecular dynamics and DFT calculations by Ye et al.[41] In has sufficient accumulated strain energy to overcome hBN's interlayer shear motion energy barrier, which supports the experimental observation that In thin films preferentially exfoliate monolayer flakes. In contrast, the total strain energies of the Pd and Au thin films differ from hBN's shear energy barrier by one and two orders of magnitude respectively. We therefore have to explore beyond the strain energy at the interface between metal and the first hBN layer to understand thicker flake exfoliation.

To further examine the exfoliation of thicker flakes, we employ an analytical model previously applied to explain thickness-selective spalling in 3D semiconductors, such as Ge, and 2D semiconductors, such as $MoS_2$.[34,45,48] In this modified Suo-Hutchinson model, strain accumulates in the bulk crystal as the result of the tensile metal thin film (Fig. S10).[45] At a given depth (termed the spalling depth), the accumulated strain energy in the thin film and bulk crystal equals the crystal's internal binding energy. During exfoliation, an adhesive handle, such as the TRT, applies an external bending moment which initiates crack formation along a preexisting crystal domain boundary or crystallographic edge.[45] The crack propagates down into the bulk crystal until it reaches the spalling depth, where it proceeds parallel to the van der Waals interface, allowing for the exfoliation of large-area flakes with uniform thickness. Within this context, spalling depth and exfoliated flake thickness are synonymous.

Using this model, we predict the relationship between residual thin film stress and spalling depth. We calculated this for bulk crystals with thicknesses of 600 nm and 100 nm respectively, and a 100 nm thick Au thin film. These bounds reflect the approximate thickness range of bulk crystals exfoliated in this work. As demonstrated with dashed lines in Fig. 4d, we observe that the model predicts higher residual stress correlating to thinner exfoliated hBN flakes, which is the same trend observed in our experimental data. This is well explained by the model as metal thin films with less initial stress require more strain to accumulate layer-by-layer in the hBN crystal in order to reach hBN's crystal binding energy.[45] Here the crystal binding energy of hBN is taken to be its van der Waals interlayer binding energy.[45,49]

However, despite the general agreement in the trend of increasing spalling depth with decreasing stress, the application of this analytical model to 2D systems has several innate limitations. The model relies on linear elastic fracture mechanics and assumes isotropic mechanical properties, which do not typically apply to layered, van der Waals materials.[50,51] Additionally, due to a lack of existing mechanical measurements on hBN (e.g. out-of-plane Young's modulus), we substituted graphene mechanical parameters into the model.[49,52,53] The precise measurement of hBN's mechanical properties and further development of a spalling model for 2D van der Waals crystals will be important to obtaining a more complete understanding of the mechanical mechanisms underpinning the metal-assisted exfoliation of hBN.

**Encapsulation of TMDC monolayer with metal-assisted exfoliated hBN**

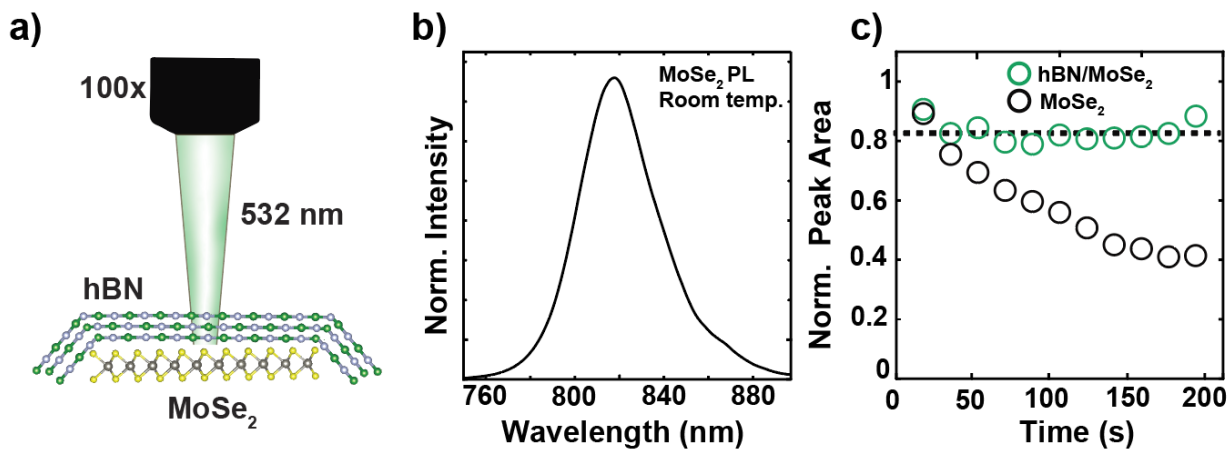

**Figure 5.** (**a**) Scheme showing PL measurement of monolayer MoSe$_2$ encapsulated with metal-exfoliated hBN. (**b**) Example PL spectrum of monolayer MoSe$_2$ (**c**) Photoluminescence peak intensity for both encapsulated and exposed monolayer MoSe$_2$, under intense 532 nm cw illumination over time.

Lastly, we use the metal-assisted method to directly exfoliate macroscopic scale hBN flakes on TMDC monolayers as encapsulating layers, without the need of flake finding or additional polymer stamp transfer steps (see Methods). Using this technique, we encapsulate a monolayer MoSe$_2$ flake with a gold-exfoliated hBN flake (~7 layers thick) and examine its photoluminescence (PL) intensity under continuous 532 nm light exposure (Fig. 5a). An example MoSe$_2$ PL spectrum is shown in Fig 5b. The encapsulated MoSe$_2$ layer maintains a fairly constant PL peak intensity under continuous, 3.4 mW laser illumination (Fig 5c), while exposed MoSe$_2$ experiences a consistent decrease in PL peak intensity overtime, indicating photodamage. These results suggest that metal-exfoliated hBN can be used effectively to encapsulate macroscopic 2D devices and protect monolayers, maintaining their optoelectronic properties under ambient conditions.[54]

## Conclusion and outlook

We demonstrate the exfoliation of millimeter scale, thin layer hBN flakes with selective thickness control using various metal thin films. The exfoliated hBN flakes demonstrate high crystalline and optical quality similar to those obtained via Scotch tape exfoliation. Our investigation of the stress, binding energy, and crystallographic alignment between the metal thin films and hBN bulk crystals indicates that residual stress from metal deposition is likely the key determinant of exfoliated flake thickness. The overall trend of high residual stress corresponding to thin hBN flakes agrees well with the applied analytical spalling model, suggesting that the application of spalling methods to 2D materials may be a promising route to industrialization. However, key improvements in bulk crystal growth and a transition from manual to robotic exfoliation will be necessary. Overall, these high-quality hBN flakes with large-area and selective thickness can support a wide range of experimental applications, which have been historically limited by flake size or thickness restrictions. This includes many ultrafast spectroscopic techniques, such as time-resolved X-ray scattering, ultrafast electron diffraction, terahertz or high harmonic generation spectroscopy, where the large effective probe sizes place constraints on sample flakes.[5,55] Additionally, this exfoliation technique is well suited to applications where specific flake thicknesses are required, such as atomically thin dielectric intermediary layers in 2D devices or artificially stacked ferroelectric hBN samples.[9,56] Lastly, the underlying strain-dependent mechanism provides a new tuning parameter in 2D material exfoliation and production. The development of similar macroscopic scale exfoliation methods will open up possibilities for the future commercialization and mass production of high-quality, single crystalline hBN-based 2D devices.

## Methods

Sample Preparation

*Thin-layer Hexagonal Boron Nitride*

Large-area, single-crystalline hBN flakes were prepared via layer-engineer exfoliation. In this method, 100 nm of the desired metal was evaporated on the surface of a freshly cleaved bulk hBN crystal (2D Semiconductor) at a rate of 0.5 Å/s using a high vacuum e-beam evaporator (LAB18 Kurt J. Lesker). The chamber pressure was maintained between $10^{-8}$ and $10^{-6}$ torr. A 10 wt% solution of Polyvinyl Pyrrolidone (PVP, M.W. 40000, Fischer) in 1:1 solution of EtOH (>95%, Fischer) and acetonitrile (99.95%, Fischer) was prepared and spin-coated onto the metal/hBN surface at 400 rpm for 2 min. Next, Thermal release tape (TRT, semiconductor equipment Corp., 90 °C release temperature) was gently pressed to the PVP/metal/hBN surface and peeled upwards to exfoliate hBN. The TRT/PVP/metal/flake system was then transferred onto the desired substrate (285 nm $SiO_2$/Si, NOVA). The system was then placed in water and heated to ~97 °C. The water environment served two key purposes. First, it promoted the dissolution of the PVP layer, helping the TRT efficiently separate from the metal surface. Second, it allowed for more even heating and therefore a more even release of the TRT and a higher overall yield of thin layer hBN. Once the TRT had been released, the samples underwent three 30 min rinses in deionized water to remove any residual PVP.

The metals were then etched using the following etchants at room temperature. Pd was etched in aqua regia for 15 min. In was etched in aqua regia for 5 min. Au was etched for 8 min in a KI/I$^-$ solution prepared by dissolving 2.5g $I_2$ (Iodine, 99.8%, Spectrum Chemical) and 10 g KI

(potassium iodide, 99.5%, Baker Chemical) in 100 ml DI water.[31,57] Both Ni and Ti were etched in 36.5% HCl (36.5%, Millipore), for 12 min and 50 min respectively.[57] Cu was etched in 51% HNO$_3$ (68%, Fischer) for 5 min each.[57] Following etching, the samples underwent three consecutive, 30 min water rinses and were briefly washed with IPA (99.5%, Fischer) and dried.

*Encapsulated molybdenum diselenide*

The monolayer MoSe$_2$ was exfoliated from bulk MoSe$_2$ crystals (HQgraphene Inc) using the macroscopic gold tape exfoliation method reported in Liu et al.[31] (See Supporting Information for details) The monolayer was transferred onto a 285 nm SiO$_2$/Si substrate. Next, a ~7L hBN flake was exfoliated with gold and thermally released onto the MoSe$_2$ sample. The gold layer on the hBN flake was then etched in a gentle KI/I$^-$ solution to yield hBN encapsulated MoSe$_2$.

Characterization

*Atomic force microscopy*

A Park NX-10 atomic force microscope was used in non-contact mode to obtain line profiles of all reported hBN flakes. Once flake step heights were obtained in a topography scan, a nominal atomic layer thickness of 0.33 nm and substrate-flake spacing of 0.3 nm were used to calculate the number of layers.[58] This procedure was performed for nearly all exfoliated flakes from each metal to determine metal-specific flake thickness selectivity. The flake thickness determined via AFM was also verified by Raman spectroscopy.

*Raman spectroscopy*

All Raman spectra of hBN's E$_{2g}$ mode were obtained using a Horiba Labram HR Evolution Raman System with a confocal geometry. Triplicate scans were collected for each location using a 532 nm with laser power of 4.92 mW and an integration time of 30 s.

*Photoluminescence*

All PL spectra of both exposed and hBN encapsulated MoSe$_2$ were obtained using the same Horiba Labram HR Evolution instrument as the Raman spectra. Sequential PL spectra were taken using a 532 nm laser at a power of 3.36 mW and an integration time of 15 s to observe photoinduced sample damage. For each spectrum, the PL peak at ~1.5 eV was numerically integrated and its integrated area examined as a function of time for both exposed and encapsulated MoSe$_2$ samples.

*Second harmonic generation*

SHG polarization scans were used to verify the quality and singly crystallinity of the exfoliated hBN. The measurements were performed on a microscope (Nikon Plan Fluor 40x objective) at room temperature. The SHG pump (1030 nm) was generated from a femtosecond laser (NKT Origami Onefive 10, pulse duration < 200 fs), and reflected off a 820 nm short-pass dichroic mirror, before being focused onto the sample by the 40x objective. The SHG signal generated by the sample is then transmitted through the 820 nm short-pass dichroic mirror and a 600nm shortpass filter before being collected by an EMCCD (Andor iXon Ultra). The signal was amplified and detected with EM gain up to 300. The polarization for both the input laser and the collection path are rotated simultaneously relative to the sample, by rotating a half waveplate between the objective and the dichroic mirror, intersecting both the excitation and collection path. Therefore, the polarization response of SHG is equal to rotating the sample, giving rise to a 6-fold symmetry.

*Selected area electron diffraction*

All SAED patterns were obtained using an FEI Tecnai G2 S-TWIN TEM. Metal/hBN samples were exfoliated using the procedure described above and transferred onto 30 nm thick Silicon Nitride membrane window TEM grids. The TEM was operated at 200 kV using a Gatan SC200 camera.

*Powder X-ray diffraction*

PXRD measurements under ambient conditions were performed using a Bruker D8 Advance diffractometer equipped with a Cu anode (K$\alpha_1$ = 1.54060 Å, K$\alpha_2$ = 1.54443 Å, K$\alpha_2$/K$\alpha_1$ = 0.50000), a divergence slit with a nickel filter, and a LYNXEYE 1D detector.

*Thin-film stress measurements*

Process induced stress due to direct metal deposition was characterized using a Flexus 2320 surface profilometer.[43] Circular coverslips (170 μm thin and 22 mm in diameter) were prepared with a 5 nm Ti-coating to improve reflectivity. The curvature of the coverslips was measured before and after the various 100 nm thick metal depositions and the Stoney equation used to calculate the induced thin film stress.[59]

*Ab initio calculation*

All the first-principle calculations are performed using the plane-wave pseudopotential density function theory (DFT) package Quantum Espresso.[60] Optimized, norm-conserving pseudopotentials generated by OPIUM package are used with the PBE exchange-correlation functional.[61–63] A cut-off energy of 50 Ryd of the plane-wave has been used to obtain a converged total energy and charge density. DFT-D3 vdW is believed to yield a better agreement for the metal lattices with the experimental results, which is applied to relax the cells and the atomic positions. To calculate the binding energy between metal surface and hBN, a slab model is set up with at least 5 layers of metal atoms and 5 layers of hBN sheets. A vacuum thickness around 14 Å is inserted to avoid the interaction with its image. To avoid the potential of dipole field to the image cell, the dipole correction is implemented. The binding energy (per BN formula unit) is computed as: $\gamma = E_{\text{metal-hBN}} - (E_{\text{metal}} + E_{\text{hBN}})$, where $E_{\text{metal-hBN}}$ is the total energy of the heterostructure, $E_{\text{metal}}$ is the total energy of the substrate metal only, and $E_{\text{hBN}}$ is the total energy of the hBN layers. When adsorbing the hBN layers into metal surfaces, the lattice of hBN is adjusted to be consistent to the metal surfaces (see Supporting Information and Fig S11).

# Acknowledgements

**Funding:** F.L. acknowledges the support of the Terman Fellowship and startup grant from Department of Chemistry at Stanford University. A.M.G. acknowledges support of the National Science Foundation Graduate Research fellowship (NSF-GRFP) under Grant No. DGE-2146755 and the John Stauffer Graduate Fellowship. F.P. and J.A.D. acknowledge support from U.S. Department of Energy, Office of Science, National Quantum Information Science Research Centers. J.S. acknowledges support from the U.S. Department of Energy, Office of Science, National Quantum Information Science Research Centers. J. H. acknowledges fellowship support from NTT Research. Y.S. acknowledges support from the Department of Energy, Office of Science, Basic Energy Sciences, Materials Science and Engineering Division, under Contract DE-AC02-76SF00515. J.L. acknowledges a Stanford Interdisciplinary Graduate Fellowship. The second-harmonic measurements were supported by Q-NEXT Quantum Center, a U.S. Department of Energy (DOE), Office of Science, National Quantum Information Science Research Center and by the Gordon and Betty Moore Foundation's EPiQS Initiative through grant number GBMF9462. Metal depositions, stress testing, Raman spectroscopy, AFM, and TEM measurements were performed at the Stanford Nano Shared Facilities (SNSF) and Stanford Nanofabrication Facility (SNF), both supported by the National Science Foundation under award ECCS-2026822.
**Supporting:** The authors thank Prof. Hemamala Karunadasa and Prof. Aaron Lindenberg for key supporting discussions of hBN applications and characterization methods, and Dr. Mathew Mate for discussions on mechanical modeling.

**Author contributions:** F.L. conceived this work. A.M.G. performed all characterizations with the exception of the XRD (performed by J.L.) and single AFM topographical scan in Fig. 2 (performed by A.S.). A.M.G. and H.Z. developed metal etching procedures. SHG measurements were performed by A.M.G. with the support of T.H. and J.H. F.Z. performed all DFT calculations. Additionally, S.W. conceived of the residual stress measurements and participated in key discussions surrounding mechanical exfoliation. J.S., Y.S., and F.P. contributed to data interpretation and analysis. J.A.D., A.M.G. and F.L. wrote the manuscript with input from all authors. Lastly, all authors contributed to the discussion and interpretation of the resulting data.
**Competing interests:** The authors are not aware of any competing interests with this work. **Data and materials availability:** All data required to evaluate the conclusions in the paper are available in the main text or supplementary materials.